\documentclass[
12pt,
preprint,preprintnumbers,nofootinbib,
groupedaddress,superscriptaddress,amsmath,amssymb]{revtex4}
\usepackage{graphicx}
\usepackage{dcolumn}
\usepackage{bm}
\usepackage{amssymb}
\usepackage{amsmath}
\usepackage{epsfig}    
\usepackage{color}
\usepackage{slashed}
\usepackage{hhline}

\def\be{\begin{equation}}
\def\ee{\end{equation}}
\newcommand{\bea}{\begin{eqnarray}}
\newcommand{\eea}{\end{eqnarray}}
\newcommand{\nn}{\nonumber}

\numberwithin{equation}{section}

\begin{document}

{\begin{flushright}{KIAS-P17069}
\end{flushright}}

\title{Hidden $U(1)$ gauge symmetry realizing a neutrinophilic two-Higgs-doublet model with dark matter}
%

\author{Takaaki Nomura}
\email{nomura@kias.re.kr}
\affiliation{School of Physics, KIAS, Seoul 02455, Korea}

\author{Hiroshi Okada}
\email{macokada3hiroshi@cts.nthu.edu.tw}
\affiliation{Physics Division, National Center for Theoretical Sciences, Hsinchu, Taiwan 300}

\date{\today}

\begin{abstract}
We propose a neutrinophilic two Higgs doublet model with hidden local $U(1)$ symmetry, where active neutrinos are Dirac type, and a fermionic DM candidate is naturally induced as a result of remnant symmetry even after the spontaneous symmetry breaking. In addition, a physical Goldstone boson is arisen as a consequence of two types of gauge singlet bosons and contributes to the DM phenomenologies as well as additional neutral gauge boson. Then we will analyze the relic density of DM within the safe range of direct detection searches, and show the allowed region of dark matter mass.
\end{abstract}
\maketitle
\newpage

\section{Introduction}
Neutrinophilic two Higgs doublet model (NTHDM)~\cite{Davidson:2009ha, Wang:2006jy, Baek:2016wml, Machado:2015sha} is one of the appropriate explanations to relax the neutrino Yukawa coupling where one of Higgs doublets has only the neutrino Yukawa interaction and develops a tiny vacuum expectation value (VEV) to generate the neutrino masses.
In order to discriminate neutrinophilic Higgs doublet from the standard model (SM) like Higgs doublet, one usually imposes an additional symmetry such as global and/or gauged one~\cite{Montero:2007cd, Ma:2014qra, Singirala:2017see, Nomura:2017vzp, Nomura:2017jxb, DeRomeri:2017oxa, Bertuzzo:2017sbj, Campos:2017dgc},
and this symmetry often plays a role in assuring stability of a dark matter candidate (DM).

 We can construct a NTHDM with extra $U(1)$ gauge symmetry assigning its charge to right-handed neutrinos and one Higgs-doublet so that this Higgs doublet only has Yukawa couplings associated with right-handed neutrino and lepton doublets. In such a case, other SM fermions would be required to have extra $U(1)$ charges for anomaly cancellation as in the $U(1)_{B-L}$ model.  Alternatively we find that we can cancel gauge anomaly among only SM singlet fermions adding extra fermions in addition to right-handed neutrinos and extra $U(1)$ gauge symmetry is a {\it hidden gauge symmery}. As a result of the gauge symmetry, the lightest extra fermions is stable and can be a good DM candidate.  

In this paper, we introduce a local hidden $U(1)$ symmetry ($U(1)_H$), and neutrino masses are Dirac type~\cite{Nomura:2017jxb} induced by the VEV of neutrinophilic Higgs doublet which has $U(1)_H$ charge.
After spontaneous symmetry breaking, a fermionic DM candidate arises as a result of remnant symmetry.
Simultaneously a physical Goldstone boson (GB) can contribute to the DM phenomenologies as well as additional neutral gauge boson, as a result of introducing two type of
gauge singlet bosons that break $U(1)_H$. We then show the observed relic density of DM can be explained either by GB interactions or $Z'$ interactions.

This paper is organized as follows.
In Sec.~II, we show our model, 
and formulate the  boson sector, fermion sector, and dark matter sector.
Then we analyze DM through the relic density and discuss the allowed region in terms of DM mass.  
Finally We conclude and discuss in Sec.~III.


 \begin{widetext}
\begin{center} 
\begin{table}[t]
\begin{tabular}{|c||c|c|c|c||c|c|c|c|c|}\hline\hline  
Fields & ~$\Phi$~ & ~$H$~ & ~$\varphi$~& ~$\varphi'$~ &~$L_{L_a}$~ & ~$e_{R_a}$~ & ~$N_{R_i}$~ & ~$N_{R_3}$~ & ~$\nu_{R_a}$~ 
\\\hline 
 $SU(2)_L$ & $\bm{2}$  & $\bm{2}$  & $\bm{1}$ & $\bm{1}$ & $\bm{2}$  & $\bm{1}$ & $\bm{1}$& $\bm{1}$& $\bm{1}$   \\\hline 
$U(1)_Y$ & $\frac12$ & $\frac12$  & $0$ & $0$  & $-\frac12$ & $-1$ & $0$ & $0$ & $0$    \\\hline
 $U(1)_H$ & $1$ & $0$  & $1$ & $8$   & $0$   & $0$  & $-4$   & $5$  & $1$    \\\hline
\end{tabular}
\caption{Field contents of bosons and fermions
and their charge assignments under $SU(2)_L\times U(1)_Y\times U(1)_{H}$ in the lepton sector, where $a=1-3$ and $i=1,2$ are flavor indices.}
\label{tab:1}
\end{table}
\end{center}
\end{widetext}

\section{ Model setup and phenomenologies}
First of all, we introduce a $U(1)_H$ hidden gauge symmetry and add six right-handed neutral fermions $(N_{R_i},N_{R_3})$ and $\nu_{R_a}$ with $i=1,2$ and a=1-3 which are charged under the new gauge symmetry.
As we discuss below, gauge anomalies are canceled among these additional fermions and  active neutrinos are Dirac type with right-handed neutrinos $\nu_{R_a}$. 
In scalar sector, we introduce an isospin doublet scalar $\Phi$ which has $U(1)_H$ charge 1, two isospin singlet bosons $(\varphi,\varphi')$ with $U(1)_H$ charges $(1, 8)$. Here $H$ is expected to be the SM-like Higgs doublet field.
All the field contents and their assignments are summarized in table~\ref{tab:1}.
Then one finds the relevant Lagrangian associated with the lepton Yukawa interactions and scalar potential as follows:
\begin{align}
-{\cal L}_{Lepton} =&
y_{\ell_{a}} \bar L_{L_a} e_{R_a} H + y_{\nu_{ab}} \bar L_{L_a} \tilde\Phi \nu_{R_b}
+ y_{\varphi_i} \varphi^* \bar N_{R_i}^c N_{R_3} + y_{\varphi'_{ij}} \varphi' \bar N_{R_i}^c N_{R_j} + {\rm c.c.},\\
V =& - \mu_H^2 H^\dagger H - \mu_\Phi^2 \Phi^\dagger \Phi - \mu_\varphi^2 \varphi^\dagger \varphi - \mu_{\varphi'}^2 \varphi'^\dagger \varphi'  \nonumber \\
& + \lambda_1 (\Phi^\dagger \Phi)^2 + \lambda_2 (H^\dagger H)^2  + \lambda_\varphi (\varphi^\dagger \varphi)^2 +  \lambda_{\varphi'} (\varphi'^\dagger \varphi')^2 
+ \lambda_{3} (H^\dagger H)(\Phi^\dagger \Phi)\nonumber \\
& + \lambda_{4} (H^\dagger \Phi)(\Phi^\dagger H) + \lambda_{H \varphi} (H^\dagger H)(\varphi^\dagger \varphi) + \lambda_{H \varphi'} (H^\dagger H)(\varphi'^\dagger \varphi') 
+ \lambda_{\Phi \varphi} (\Phi^\dagger \Phi)(\varphi^\dagger \varphi) \nonumber \\ 
& + \lambda_{\Phi \varphi'} (\Phi^\dagger \Phi)(\varphi^\dagger \varphi) + \lambda_{\varphi \varphi'} (\varphi^\dagger \varphi)(\varphi'^\dagger \varphi')  - \mu_0\left[(\Phi^\dag H) \varphi +{\rm c.c.}\right] 
,
\label{eq:lag-lep}
\end{align}
where $\tilde \Phi \equiv (i \sigma_2) \Phi^*$ with $\sigma_2$ being the second Pauli matrix, $a$ runs over $1$ to $3$, and $i,j$ runs over $1$ to $2$. The first term of Yukawa coupling provides the SM charged leptons masses, while the second term induces the active neutrino masses of Dirac type. 
The term $\mu_0$ plays a role in forbidding a massless Goldstone boson appearing from Higgs doublets after spontaneous gauge symmetry breaking~\cite{Baek:2016wml}. 
Note that we have $Z_2$ symmetry even after scalar fields developing VEVs where extra fermions $\{N_{R_i}, N_{R_3} \}$ are $Z_2$ odd and the other particles are $Z_2$ even at renormalizable level
\footnote{At non-renormalizable level, there exists dimension 6 operator of $\bar \nu_{R_{a}}^c N_{R_3} (\varphi')^*\varphi^2$.
We consider such a term is highly suppressed by sufficiently large cut-off scale as well as its coupling and suppose not to affect stability of DM and phenomenology.}.

{\it Anomaly cancellation:} Here we check anomaly cancellations for new gauge symmetry in the model. In our case, we need to check only $U(1)_H$ and $[U(1)_H]^3$ anomalies since all the $U(1)_H$ charged fermions are SM singlet. We then find:
\begin{align}
& U(1)_H : \quad 1+1+1-4-4+5 =0 \nonumber \\
& [U(1)_H]^3: \quad (1)^3+(1)^3+(1)^3+(-4)^3+(-4)^3+(5)^3=0.
\end{align}
Therefore our charge assignment is anomaly free.

{\it Scalar sector}:
The scalar fields are parameterized as 
\begin{align}
&H =\left[\begin{array}{c}
w^+\\
\frac{v_H + h +i z}{\sqrt2}
\end{array}\right],\ 
\Phi =\left[\begin{array}{c}
\phi^+\\
\frac{v_\phi + \phi_R + i \phi_I}{\sqrt2}
\end{array}\right],\ 
\varphi=
\frac{v_\varphi + \varphi_{R}}{\sqrt2}{e^{i \frac{ \alpha}{v_\varphi }}},\ 
\varphi'=
\frac{v_{\varphi'} + \varphi'_{R}}{\sqrt2} {e^{i \frac{ \alpha'}{v_{\varphi'} } }},
\label{component}
\end{align}
where the lightest mass eigenstate after diagonalizing the matrix in basis of $(w^\pm$, $\phi^\pm)$, which is massless, is absorbed by the SM singly-charged gauge boson $W^\pm$, and two degrees of freedom in the CP-odd boson sector  $(z, \phi_I, \alpha, \alpha')$ are also absorbed by the neutral SM gauge boson $Z$ and  $U(1)_H$ gauge boson $Z'$~\footnote{Since the structure of scalar sector is more or less the same as the one in ref.~\cite{Nomura:2017jxb}, we minimally explain properties of the scalar bosons.}; $z$ is dominantly NG boson absorbed by $Z$ and one linear combination of $\{ \alpha, \alpha' \}$ is absorbed by $Z'$ as discussed below.
The non-zero VEVs of scalar fields can be obtained from the condition $\partial V/\partial v_{H, \phi, \varphi, \varphi' } =0 $.  
Then we can simply obtain
\begin{align}
v_\varphi \simeq \sqrt{\frac{\mu_\varphi^2}{\lambda_\varphi}}, \quad v_\varphi \simeq \sqrt{\frac{\mu_{\varphi'}^2}{\lambda_{\varphi'}}}, \quad
v_H \simeq \sqrt{\frac{\mu_H^2}{\lambda_H}}, \quad v_\phi \simeq \frac{\sqrt{2} \mu_0 v_H v_\varphi }{ -2 \mu_\Phi^2 + (\lambda_{3} + \lambda_{4})v_H^2}  
\end{align} 
where we assumed couplings in the potential $\{\lambda_{H \varphi}, \lambda_{H \varphi'}, \lambda_{\Phi \varphi}, \lambda_{\Phi \varphi'}, \lambda_{\varphi \varphi'} \}$  and $v_\phi$ to be sufficiently small, 
 and we require $\{\mu_H^2, \mu_\varphi^2, \mu_{\varphi'}^2, \mu_0 \} > 0$ and $-2 \mu_\Phi^2  + (\lambda_3 + \lambda_4) v_H^2 >0 $ to make all VEVs positive. Note that $v_\phi$ is expected to be tiny in order to generate active neutrino mass which can be realized taking tiny $\mu_0$ value~\cite{Baek:2016wml,Nomura:2017jxb}. 
One thus finds that $v\equiv \sqrt{v_H^2+v_\phi^2}\sim v_H$.
The mass matrix squared of singly charged boson is diagonalized by the following mixing matrix as
\begin{align}
O
&\equiv
\left[\begin{array}{cc}
c_\beta&  s_\beta \\ 
-s_\beta & c_\beta \\ 
\end{array}\right], \quad s_\beta=\frac{v_\phi}{\sqrt{v_\phi^2+v_H^2}},
\end{align}
where we define ${\rm Diag.}(0, m^2_{H^\pm})=O m^2(w^\pm, \phi^\pm)O^T$.
Therefore we obtain
\begin{align}
 \phi^\pm \simeq H^\pm.
\end{align}
The mass of the charged Higgs boson is given by
\begin{equation}
m_{\phi^\pm}^2  \simeq  - \mu_\Phi^2 + \frac{1}{2} (\lambda_3 - \lambda_4) v_H^2.
\end{equation}
Then $\mu_\Phi^2$ is further constrained requiring $m_{\phi^\pm}^2 > 0$ in addition to condition for obtaining positive VEV of $\Phi$.
On the other hand the mass squared matrix of the CP-odd boson is in basis of $(z, \phi_I, {\alpha, \alpha'})$.
After diagnolizing the mass matrix, we obtain one massive CP-odd scalar, two NG boson absorbed by $Z$ and $Z'$ boson, and one massless physical Goldstone boson.
We can identify massive CP-odd scalar as $\phi_I$ whose mass is given by
\begin{equation}
m_{\phi_I}^2  \simeq  - \mu_\Phi^2 + \frac{1}{2} (\lambda_3 + \lambda_4) v_H^2.
\end{equation}
 In general scalar bosons $\{\phi_R, \phi_I, \phi^\pm \}$ mix with other scalar degrees of freedom which have same quantum number. However those mixings are highly suppressed in our scenario where $v_\phi$ is assumed to be tiny in realizing neutrino mass. For example, if we take $y_\nu = 10^{-6}(\sim m_e/v)$ required value of $v_\phi$ is less than $\sim 100$ KeV as $m_\nu \sim y_\nu v_\phi$, and mixing effect is roughly given by $v_\phi/m_{\rm scalar}$ which is negligibly tiny taking $m_{\rm scalar} = \mathcal{O}(100)$ GeV scale.

{The NG boson absorbed by $Z'$ and physical Goldstone boson are written in terms of linear combination of $\alpha$ and $\alpha'$ where the mixing angle is determined by relative sizes of VEVs of $\varphi$ and $\varphi'$. We then obtain NG and physical Goldstone modes denoted by $\alpha_{NG}$ and $\alpha_G$ such that~\footnote{Derivation of these states is summarized in the Appendix.} 
\begin{align}
& \alpha_{NG} = c_X \alpha + s_X \alpha', \quad \alpha_G = - s_X \alpha + c_X \alpha' , \label{eq:GandNG1} \\
& c_X \equiv \cos X = \frac{v_\varphi}{\sqrt{v_\varphi^2 + 64 v_{\varphi'}^2 }}, \quad s_X \equiv \sin X = \frac{8v_{\varphi'}}{\sqrt{v_\varphi^2 + 64 v_{\varphi'}^2 }}. \label{eq:GandNG2}
\end{align} }
Notice that the existence of this physical Goldstone boson does not cause serious problem in particle physics or cosmology since it does not couple to SM particles directly and decouples from thermal bass in early Universe.

The extra gauge boson $Z'$ obtain mass after $U(1)_H$ symmetry breaking as 
\begin{equation}
m_{Z'}^2 = g_H^2 (v_\varphi^2 + 64 v_{\varphi'}^2), \label{eq:zp-mass}
\end{equation}
where $g_H$ denotes the gauge couplings for $U(1)_H$ gauge symmetry.
 Note that we can have $Z$-$Z'$ mixing through the VEV of $\Phi$ since it has both electroweak and $U(1)_H$ charge. In our case, however, it is negligibly small due to small $v_\phi$ where mixing is suppressed by the $(v_\phi/m_{Z'})^2$ factor.

Inserting tadpole conditions, the mass matrix for CP-even boson in basis of $(h,\phi_{R},\varphi_{R},\varphi'_{R})$ with nonzero VEVs is defined by $m_R^2$. Then the mixing matrix $O_R$ to diagonalize the mass matrix is defined to be $m_{h_{a}}=O_R m_R^2 O_R^T$ and $(h,\phi_{R},\varphi_{R},\varphi'_{R})^T=O_R^T h_a$ where $m_{h_{a}}$ is diagonal  mass matrix and the mass eigenstate is $h_{a}$ ($a=1-4$). Here $h_1\equiv h_{SM}^T$ is the SM Higgs, therefore,  $m_{h_1}=$125 GeV.
{ In addition, we assume mixing among SM Higgs and other CP-even scalars are small to avoid experimental constraints for simplicity.}

\subsection{Fermion sector}
First of all, we formulate the mass matrix of the SM leptons.
The masses for charged-leptons are induce via  $y_\ell$ after symmetry breaking,
and active neutrino masses are also done via $y_\nu$  term where neutrinos are supposed to be Dirac type fermions.
Their masses are symbolized by $m_{\ell_a}\equiv v_H y_{\ell_a}/\sqrt2$ and $m_{\nu_{ab}}\equiv v_\phi y_{\nu_{ab}}/\sqrt2$.
Since the charged-lepton mass matrix is diagonal, the neutrino mixing matrix $V$ is arisen from the neutrino mass matrix squared; $(m_\nu^2)_{ab} = \sum_{c=1-3}(m_{\nu_{ac}}m_{\nu_{cb}}^\dag)$, where $V$ is measured by the neutrino oscillation data~\cite{Tortola:2012te}.
Notice here that three active neutrinos can have non-zero mass due to the rank three matrix.
In our scenario we take $y_\nu$ not to be very large such as $y_\nu \sim 10^{-6}(\sim m_e/v)$.
 Note that our right-handed neutrinos decouple from thermal bath sufficiently earlier than left-handed neutrinos since $y_\nu$ coupling is small and $Z'$ mass is heavier than electroweak scale. Thus they do not affect cosmological issues such as Big Bang nucleosynthesis.

{\it Majorana fermions}:
Then we formulate the mass matrix of exotic Majorana fermions $M_N$ in basis of $(N_{R_1},N_{R_2},N_{R_3})^T$,
which is give by 
\begin{align}
M_N&\equiv
\left[\begin{array}{ccc}
m_{11} & m_{12} & M_1 \\ 
m_{12} & m_{22} &  M_2 \\ 
M_1 & M_2  & 0 \\ 
\end{array}\right],
\end{align}
after spontaneous $U(1)$ breaking, where $m_{ij}(=m_{ji})\equiv y_{\varphi'_{ij}}v_{\varphi'}/\sqrt2$, and $M_{i}\equiv y_{\varphi_{i}}v_{\varphi}/\sqrt2$,
with $i,j=1,2$.
Then $M_N$ is diagonalized by $D(M_{\psi_1},M_{\psi_2},M_{\psi_3})= V_N M_N V_N^T$.
Thus one finds $(N_{R_1},N_{R_2},N_{R_3})^T\equiv V_N^T ({\psi_1},{\psi_2},{\psi_3})^T$, where $V_N$ is an unitary mixing matrix in general.
Here we take $N_{R_1}$ is the lightest mass eigenstate and it is stable particle due to the remnant $Z_2$ symmetry as discussed above. Thus we writhe $X_R \equiv \psi_{1}$ and $M_X \equiv M_{\psi_1}$ for our DM candidate in the following analysis.

\subsection{Dark matter} 
In this subsection we discuss a dark matter candidate; $X_R$.
{\it Firstly, we assume contribution from the Higgs mediating interaction is negligibly small and DM annihilation processes are dominated by the gauge interaction with $Z'$ and/or GB $\alpha_G$; we thus can easily avoid the constraints from direct detection searches as LUX~\cite{Akerib:2016vxi}, XENON1T~\cite{Aprile:2017iyp}, and PandaX-II~\cite{Cui:2017nnn}.} 

{\it Relic density}: We have annihilation modes with Yukawa and kinetic terms to explain the relic density of DM:
$\Omega h^2\approx 0.12$~\cite{Ade:2013zuv}, and their relevant Lagrangian in basis of mass eigenstate is found to be {
\begin{align}
-{\cal L} \supset &
\frac12 Q^X_{H} g_{H} \bar X\gamma^\mu \gamma_5 X Z'_\mu + g_{H} Q^\nu_{H} \bar\nu \gamma^\mu P_R \nu Z'_\mu 
+i \frac{\tilde M_{1\beta}}{\tilde v_{\varphi \varphi'}}  \bar X P_R \psi_\beta  \alpha_{G}+{\rm c.c.}, \nonumber \\
& + i g_{ H} Z'^\mu (\partial_\mu H^- H^+ - H^- \partial_\mu H^+) + g_H Z'^\mu (\partial_\mu \phi_I \phi_R - \phi_I \partial_\mu \phi_R), \\
\frac{\tilde M_{\alpha \beta}}{\tilde v_{\varphi \varphi'}} \equiv & \sum_{i=1,2} \frac{M_i}{v_\varphi} s_X (V_N)_{\alpha i} (V^T_N)_{3 \beta} + \sum_{i,j=1,2} \frac{m_{ij}}{v_{\varphi'}} c_X (V_N)_{\alpha i} (V_N^T)_{j \beta},
\label{eq:dmint}
\end{align}
where $M_i = y_{\varphi_i} v_\varphi/\sqrt{2}$, $m_{ij}\equiv y_{\varphi'_{ij}}v_{\varphi'}/\sqrt2$,  $Q^X_{H}\equiv -4 + 9 |V_{N_{13}}|^2$ is the DM charge of hidden symmetry, $Q^\nu_{H}=1$ is the active neutrino charge of hidden gauge symmetry. }
Notice here that we have used the unitarity of $V_N$ to derive $Q^X_{H}$;
 $\sum_{a=1}^3V^*_{N_{1a}} V^T_{N_{a1}}=1$. 
The first and second terms induce the mode of active neutrino final state via $Z'$ vector boson exchange in $s$-channel; the last two terms also provides final state containing new scalar bosons from second Higgs doublet.
On the other hand the third term induces the annihilation process where the final state is GB via the diagrams with neutral fermions in the $t$ and $u$ channels.
The relic density of DM is then given by~\cite{Griest:1990kh, Edsjo:1997bg}
\begin{align}
&\Omega h^2
\approx 
\frac{1.07\times10^9}{\sqrt{g_*(x_f)}M_{Pl} J(x_f)[{\rm GeV}]},
\label{eq:relic-deff}
\end{align}
where $g^*(x_f\approx25)$ is the degrees of freedom for relativistic particles at temperature $T_f = M_X/x_f$, $M_{Pl}\approx 1.22\times 10^{19}$ GeV,
and $J(x_f) (\equiv \int_{x_f}^\infty dx \frac{\langle \sigma v_{\rm rel}\rangle}{x^2})$ is given by~\cite{Nishiwaki:2015iqa}
\begin{align}
J(x_f)&=\int_{x_f}^\infty dx\left[ \frac{\int_{4M_X^2}^\infty ds\sqrt{s-4 M_X^2} [W_{Z'}(s)+W_{\alpha_G}(s)] K_1\left(\frac{\sqrt{s}}{M_X} x\right)}{16  M_X^5 x [K_2(x)]^2}\right],\\ 
W_{Z'}(s)
\approx &\frac{(s-4M_X^2)}{8\pi}  \left| \frac{g_{H}^2 Q_{H}^X}{s-m_{Z'}^2+i m_{Z'} \Gamma_{Z'}}\right|^2 \left(s+ \frac{1}{2} (s - 4 m_{\Phi}^2) \right),\\
 W_{\alpha_G}(s)
\simeq &
\frac{|\tilde M_{11}|^4}{64\pi \tilde{v}^4_{\varphi \varphi'}} 
\left[
(3s^2-4M_X^4) \left( \frac{\pi}{2sM_X^2}\sqrt{\frac{M_X^4}{4sM_X^2-s^2}} - \frac{\tan^{-1}\left[\frac{s-2M_X^2}{\sqrt{s(4M_X^2-s)}}\right]}{s^{3/2}\sqrt{4M_X^2-s}} \right)
-4\right],  
\label{eq:relic-deff}
\end{align}
where we implicitly impose the kinematical constraint above, take degenerate $H^\pm(\phi_{R,I})$ mass as $m_\Phi$, and $XX \to Z' Z'$ process is omitted here for simplicity.
{Here $Z'$ can decay into $\nu_R \bar \nu_R$, $\psi_\alpha \psi_\alpha$ and $H^+ H^-(\phi_R \phi_I)$ if kinematically allowed.
The decay width of $Z'$, which consists of $\Gamma_{Z'}\equiv \Gamma_{Z' \to \nu_R \bar\nu_R}+\Gamma_{Z' \to X\bar X} + \Gamma_{Z' \to H^+H^-} + \Gamma_{Z' \to \phi_R \phi_I}$, 
is given by
\begin{align}
\Gamma_{Z' \to \nu_R \bar\nu_R} &= \frac{g_{H}^2m_{Z'}}{8 \pi} ,
\\
 \Gamma_{Z' \to X\bar X} &= \frac{g_{H}^2m_{Z'}}{96 \pi}  |Q_{H}^X|^2 \left( 1 - \frac{4 M_X^2}{m_{Z'}^2} \right)^{3/2}, \\
 \Gamma_{Z' \to H^+ H^- (\phi_R \phi_I)} & = \frac{g_H^2}{48 \pi} m_{Z'} \left( 1 - \frac{4 m_{\Phi}^2}{m_{Z'}^2} \right)^{\frac32},
\end{align}
where we assume masses of $\psi_{2,3}$ are heavier than $m_{Z'}/2$, 
and $N_c^f$ is color factor.
Remind here that $Z'$ mass is given by $m_{Z'} = g_{H} \sqrt{v^2_\varphi+(8 v_{\varphi'})^2}$ in Eq.~(\ref{eq:zp-mass}). }

\begin{figure}[t]
\centering
\includegraphics[width=10cm]{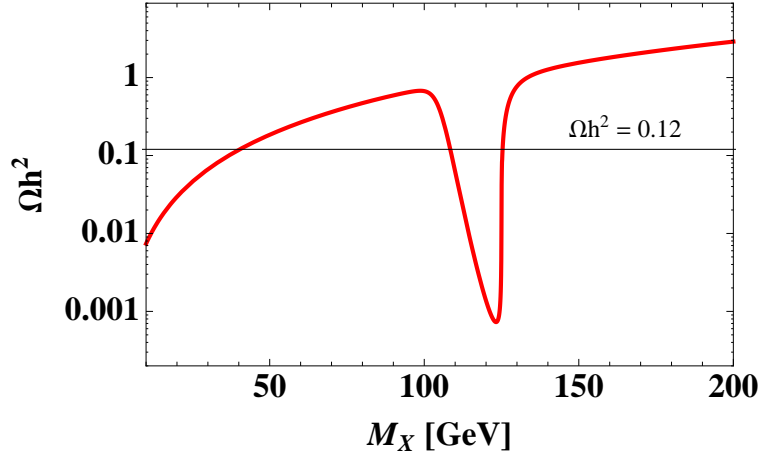}
\caption{The correlation between $M_X$ and $\Omega h^2$, where the horizontal black line is 0.12. The other parameters are fixed as given in Eq.~(\ref{eq:setting}).
}
\label{fig:relic}
\end{figure}
In fig.~\ref{fig:relic}, we show the relic density in terms of $M_X$, where we fix the following parameters\footnote{In principle, one has to derive this mixing and their masses by diagonalizing $M_N$ in the neutral fermions. But here we expect any values can be taken, since all the mass parameters except the DM mass and its mixing are free.}:
\begin{align}
 & g_{H}=0.05,\ |V_{N_{13}}|=0.1,\ m_{Z'}=250\ {\rm GeV}, \ \tilde{v}_{\varphi \varphi'}=100 \ {\rm GeV}, \nn\\
&  |\tilde M_{11}|=20 \ {\rm GeV}, \ M_{\psi_2}=500 \ {\rm GeV}, \ M_{\psi_3}=1000 \ {\rm GeV}, \ m_\Phi = 500 \ {\rm GeV}.
\label{eq:setting}
 \end{align}
The figure suggests the following allowed range for $0.05\le g_{H}$;
\begin{align}   
40\ {\rm GeV}\lesssim M_X\lesssim 110\ {\rm GeV}, \ {\rm and}\  125\ {\rm GeV}\lesssim  M_X,
\end{align}
while for $g_{H}\le 0.05$;
\begin{align}   
M_X\lesssim 40\ {\rm GeV}, \ {\rm and}\  110\ {\rm GeV}\lesssim  M_X\lesssim 125\ {\rm GeV},
\end{align}
when all the parameters except $g_{H}$ are fixed and this region indicate that observed relic density is obtained around resonant point $M_X\sim m_{Z'}/2$ where $m_{Z'}$ is proportional to $g_H$. 
\begin{figure}[t]
\centering
\includegraphics[width=10cm]{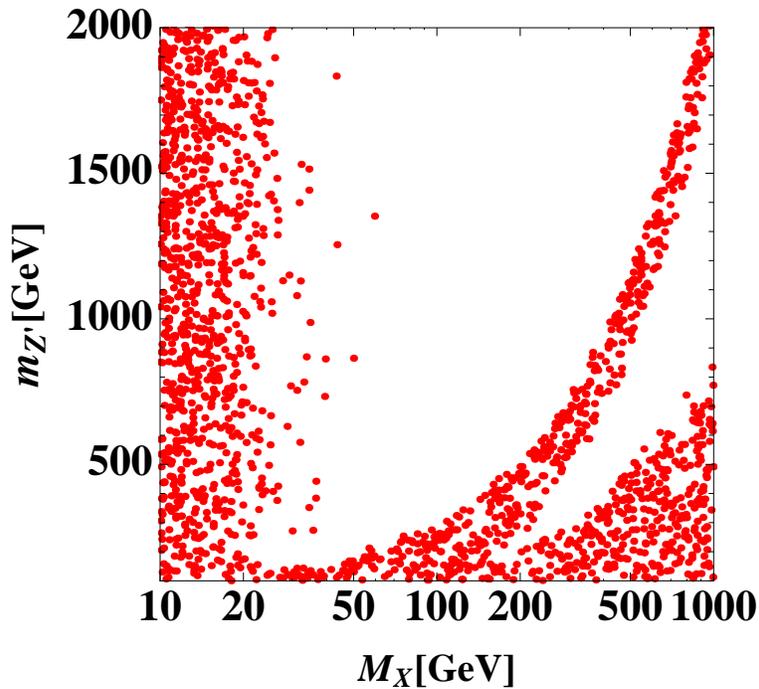}
\caption{The correlation between $M_X$ and $m_{Z'}$ when the estimated relic density is $0.11 < \Omega h^2 < 0.13 $.
The other parameters are fixed as given in Eq.~(\ref{eq:setting2}).}
\label{fig:relic2}
\end{figure}
{Here we search for parameter region satisfying observed relic density in general where we apply {\tt micrOMEGAs 4.3.5}~\cite{Belanger:2014vza} to estimate the annihilation cross sections.
Note that $XX \to Z' Z'$ process is also included in following analysis.
Then we scan parameter region as follows: 
\begin{align}
& M_X \in [10, 1000] \ {\rm GeV}, \ m_{Z'} \in [100, 2000] \ {\rm GeV}, \ \frac{\tilde{M}_{\alpha \beta}}{\tilde{v}_{\varphi \varphi'}} \in [0.025,0.4],   \nonumber \\
& g_H \in [0.05, 0.4], \ V_{13} \in [0.1, 1/\sqrt{2}], \ M_{\psi_{1,2}} \in [M_X, 1500] \ {\rm GeV}, \ m_\Phi \in [100, 300] \ {\rm GeV}.
\label{eq:setting2}
\end{align}
In fig.~\ref{fig:relic2}, we also show the parameter points on $M_X$-$m_{Z'}$ plane which give relic density $0.11 < \Omega h^2 < 0.13$ fixing the other parameters as given in Eq.~(\ref{eq:setting2}).
We find that several specific region can explain relic density of DM: (1) in light $M_X$ region $XX \to \alpha_G \alpha_G$ process is dominant one and insensitive to $m_{Z'}$, 
(2) the line shaped region indicate $m_{Z'} \sim 2 M_X$ in which relic density is explained with resonant effect, (3) in heavy $M_X$ region, relic density can be explained by $XX \to Z' Z'$ process with relevant value of $g_H$.
In addition, we show DM annihilation cross section at the current universe for parameter region giving right relic density. 
The cross section is suppressed for $\alpha_G \alpha_G$ and $Z' Z'$ modes while it can be $\sim 10^{-26}$cm$^3$/s for $\nu_R \bar \nu_R$ and $H^+H^-(\phi_R \phi_I)$ modes.
Since ratio of $H^+H^-$ mode is around $10 \%$ in the latter case our scenario is safe from constraints of indirect detection experiments, and it would be tested in future measurements of gamma-ray and neutrino flux from DM annihilation. 
}

%

\begin{figure}[t]
\centering
\includegraphics[width=10cm]{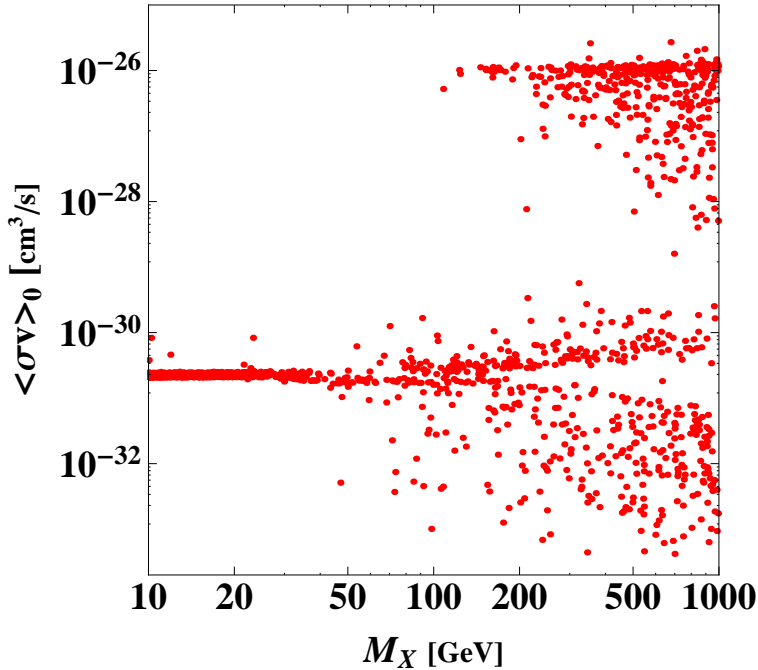}
\caption{The DM annihilation cross section at the current universe for the parameter region which provide correct relic density.}
\label{fig:CS}
\end{figure}

\section{Conclusion}
We have proposed a neutrinophilic two Higgs doublet model with hidden local $U(1)_H$ symmetry introducing right-handed neutrinos and exotic SM singlet fermions for anomaly cancellation.
The active neutrino masses are Dirac type induced by the tiny VEV of neutrinophilic Higgs doublet whose interaction to other SM fermions are forbidden by the $U(1)_H$ symmetry.
We then formulated the boson and fermion sector where a fermionic DM candidate naturally arises as the lightest mass eigenstate of exotic fermion since it is stable due to a remnant symmetry even after the spontaneous symmetry breaking.
Then the DM candidate interacts with active neutrinos by exchanging $Z'$ boson from $U(1)_H$.
Moreover, a physical GB is induced as a consequence of two types of gauge singlet scalar fields and contributes to the DM annihilation processes determining the relic density.
Then we have analyzed the relic density of DM, within the safe range of direct detection searches,
and found another allowed range with lighter DM mass that directly comes from the contribution of GB mode
in addition to the resonant allowed range via $Z'$ boson.   

\section*{Acknowledgments}
\vspace{0.5cm}
H. O. is sincerely grateful for the KIAS member and all around.

\appendix
\section{ Deriving $\alpha_G$ and $\alpha_{NG}$ in Eqs.~(\ref{eq:GandNG1}) and (\ref{eq:GandNG2})}

Here we derive NG boson $\alpha_{NG}$ and physical Goldstone boson $\alpha_G$ from $\varphi$ and $\varphi'$ expressed as in Eq.~(\ref{component}) where mixing between $\phi_I$ is ignored assuming tiny $v_\phi$.
The covariant derivative of $\varphi(\varphi')$ is given by
\begin{equation}
D_\mu \varphi (\varphi') = e^{i \frac{\alpha (\alpha')}{v_{\varphi(\varphi')}}} \left( \partial_\mu + i \frac{1}{v_{\varphi(\varphi')}} \partial_\mu \alpha (\alpha') - i g_H Q_{\varphi(\varphi')} Z'_\mu \right) r_\varphi(r_{\varphi'}),
\end{equation}
where $Q_{\varphi(\varphi')} = 1(8)$ is $U(1)_H$ charge of $\varphi(\varphi')$ and $r_{\varphi(\varphi')} = [v_{\varphi(\varphi')} + \varphi_R(\varphi'_R)]/\sqrt{2}$.
We then have 
\begin{align}
\mathcal{L}_{\rm kinetic} = & (D_\mu \varphi)^\dagger (D^\mu \varphi) + (D_\mu \varphi')^\dagger (D^\mu \varphi') \nonumber \\
 =& \frac{1}{2} \partial_\mu \varphi_R \partial^\mu \varphi_R + \frac{1}{2} \partial_\mu \varphi'_R \partial^\mu \varphi'_R  \nonumber \\
& + \frac{1}{2} (v_\varphi^2 + 2 v_\varphi \varphi_R + \varphi_R^2) \left( \frac{1}{v_\varphi^2} \partial_\mu \alpha \partial^\mu \alpha - \frac{2 g_H Q_\varphi}{v_\varphi} \partial_\mu \alpha Z'^\mu + g_H^2 Q_\varphi^2 Z'_\mu Z'^\mu \right) \nonumber \\
& + \frac{1}{2} (v_{\varphi'}^2 + 2 v_{\varphi'} \varphi'_R + \varphi'^2_R) \left( \frac{1}{v_{\varphi'}^2} \partial_\mu \alpha' \partial^\mu \alpha' - \frac{2 g_H Q_{\varphi'}}{v_{\varphi'}} \partial_\mu \alpha' Z'^\mu + g_H^2 Q_{\varphi'}^2 Z'_\mu Z'^\mu \right).
\label{eq:kinetic}
\end{align}
Here we add gauge fixing term;
\begin{align}
&\mathcal{L}_G = -\frac{1}{2} G^2, \nonumber \\
& G = \frac{1}{\sqrt \xi} (\partial_\mu Z'^\mu + \xi g_H Q_{\varphi} v_\varphi \alpha + \xi g_H Q_{\varphi'} v_{\varphi'} \alpha' ),
\label{eq:Gfixing}
\end{align}
where $\xi$ is a gauge fixing parameter. 
Combining Eq.~(\ref{eq:kinetic}) and (\ref{eq:Gfixing}), we obtain mass terms for $Z'$ and $\alpha(\alpha')$ such that
\begin{align}
\mathcal{L}_M =& \frac{1}{2} g_H^2 (Q_{\varphi}^2 v_{\varphi}^2 + Q_{\varphi'}^2 v_{\varphi'}^2) Z'_\mu Z'^\mu \nonumber \\
& - \frac{1}{2} \xi g_H^2 (Q_{\varphi}^2 v_{\varphi}^2 + Q_{\varphi'}^2 v_{\varphi'}^2)  
\left[ \frac{Q_{\varphi} v_{\varphi}}{\sqrt{ Q_{\varphi}^2 v_{\varphi}^2+ Q_{\varphi'}^2 v_{\varphi'}^2 }} \alpha + \frac{Q_{\varphi'} v_{\varphi'}}{\sqrt{Q_{\varphi}^2 v_{\varphi}^2+ Q_{\varphi'}^2 v_{\varphi'}^2}} \alpha' \right]^2.
\end{align}
Thus the second term correspond to gauge dependent mass term for NG boson, and physical Goldstone boson state is orthogonal to NG boson one.
Therefore the $\alpha_{NG}$ and $\alpha_G$ are given as Eqs.~(\ref{eq:GandNG1}) and (\ref{eq:GandNG2}).

\end{document}